\documentstyle[11pt,paspconf]{article}

\input epsf.tex

\begin{document}

\title{ On Formation of Disk Galaxies from Spherical Primordial Fluctuations}

\author{Vladimir Avila-Reese and Claudio Firmani}
\affil{Instituto de Astronom\'{\i}a, U.N.A.M., \\
Apdo. Postal 70-264, 04510 M\'exico D.F., Mexico}

\affil{Centro de Instrumentos, U.N.A.M., \\
Apdo. Postal 70-186, 04510 M\'exico D.F., Mexico }

% The abstract is entered in a LaTeX "environment", designated with paired
% \begin{abstract} -- \end{abstract} commands.  Other environments are
% identified by the name in the curly braces.

% Poster authors ONLY may omit the abstract in order to gain a little
% more page space for the text of the poster.

\begin{abstract}
 The spherical gravitational collapse and virialization of arbitrary density fluctuations in an expanding universe is studied. In the context of the standard cosmological model and the peak $ansatz$, disk galaxies are supposed to be the product of "extended" gravitational collapses of regions surrounding local-maxima of the density fluctuation field. The substructure over the main profile (merging) is considered as a second-order phenomenon. The range and distribution of possible mass growth histories or accretion regimes for a given present-day mass is estimated and used as the initial condition for galaxy formation. Considering the effect of dark halo contraction due to disk formation, the final rotation curves are calculated. If the structural properties of dark halos are important in establishing the Hubble sequence's properties of visible
galaxies, then this sequence is defined not only by the mass, but among other possible parameters, by the accretion regime.
\end{abstract}

% Keywords should be included, but they are not printed in the hardcopy.


% That's it for the front matter.  On to the main body of the paper.
% We'll only put in tutorial remarks at the beginning of each section
% so you can see entire sections together.

\section{Introduction}

 The Alpine mountains here in Sesto are a vivid representation of what could
have been the primordial density fluctuation field: peaks, mountainsides,
valleys; some distributed in a superposed fashion, others isolated and so
on.{} If we accept the gravitational paradigm (for alternatives see Babul $%
et $ $al.$ 1994) and the peak density $ansatz$ then the cosmic structures
that one observes today should have emerged from the peaks of this initial
density fluctuation field. Although some N-body experiments have shown that
the peaks can be disrupted before collapsing it is still precipitous to
make conclusions when in fact one is using inaccurate peak statistics
(Manrique \&\ Salvador-Sole 1996 and references therein). Anyway, in the
case of disk galaxies the peak $ansatz$ is more suitable because they are
typically located in less dense environments where the shear caused by the
surrounding fluctuations is less important. The peaks from which they
emerge, not being enhanced by larger scale fluctuations, should have been
high, therefore the gravitational potential of the matter collapsing around
them would tend to be spherically symmetric.
 From the gravitational collapse of the regions around peaks arise the dark
matter halos, where subsequently the baryon matter cools and falls to the
center forming the visible galaxies. A link (if any) between the present-day
observable properties of galaxies and the initial conditions can be
established only by taking into account the intermediate processes of the
formation and evolution of galaxies. Our presentations, this paper and
Firmani $et$ $al.1996$ (FAH), go in this direction.

\section{Spherical gravitational collapse}

The spherical gravitational collapse model (Gunn \&\ Gott 1972, Peebles
1980) allows us to calculate the imaginary structure one would obtain by
freezing all collapsing shells at their turn around radius:

\begin{equation}
\rho _m(r_m)=\frac{9\pi ^2}{16(1+\alpha )}\frac 1{6\pi Gt_m^2}\propto \frac{%
r_m^{-\frac{3\alpha }{1+\alpha }}}{1+\alpha }  \label{eq1}
\end{equation}
where $\alpha (x)\equiv -\frac{d\ln \Delta _0(x)}{d\ln x}$ is the local
slope of the cumulative density profile $\Delta _0(x)\propto x^{-\alpha
(x)}, $ and the proportionality in eq. 1 is valid for self-similar collapse.
If the final virialized radius $r_v$ scales with $r_m$, that is $r_v=Fr_m$
with $F=const$ for a given $\alpha ,$ then the density profile of the halo
will be

\begin{equation}
\rho _v(r_v)\propto \frac{r_v^{-\frac{3\alpha }{1+\alpha }}}{1+\alpha }
\label{eq2}
\end{equation}

Because our approach will tend to tune the present-day disk
galaxy properties to their initial conditions ($FAH$), it is more
convenient to express these initial conditions through the mass growth
history ($MGH$) back in time, rather than through an initial density
profile (both are equivalent from a conceptual point of view). Here we
parametrize the $MGH$ by a power-law:

\begin{equation}
M(t)=M_0\left( \frac t{t_0}\right) ^\gamma  \label{eq3}
\end{equation}
where $\gamma (t)$, the accretion parameter, is constant in the case
of self-similar collapse, and M$_0$ is the mass at the present epoch $t_0$.

It is straighforward to calculate eq. 2 for this case:

\begin{equation}
\rho _v(r_v)\simeq 6\cdot 10^5\left( \frac F{0.5}\right) ^2\rho
_0r_v^{-\frac 6{2+\gamma }}  \label{eq4}
\end{equation}
where $F$ is the collapse factor and $\rho _0$ is the present-day critical
density (only an Einstein-de Sitter model is considered here, and $%
H_0=50Kms^{-1}Mpc^{-1}$). The rotation curve will be

\begin{equation}
V(r_v)\simeq 27.3\left( \frac F{0.5}\right) ^{-1/2}\frac{M_{10}^{\frac
1{2+\gamma }}}{\left[ 58\left( \frac{0.5}F\right) \right] ^{\frac{\gamma -1}{%
2+\gamma }}}r_v^{-\frac{\gamma -1}{2+\gamma }}Km/s \label{eq5}
\end{equation}
where $M_{10}$ is $M_0$ in units of 10$^{10}M_{\odot }.$ Some interesting
features can be readily seen from these simple analytical results. When $%
\gamma \approx 1$ (the high accretion regime) the asymptotic rotation curve
is flat, while if $\gamma \ga 0$(the low accretion regime) the halo is a
very concentrated structure.

\begin{figure}
\vspace{8.0cm}
\includegraphics{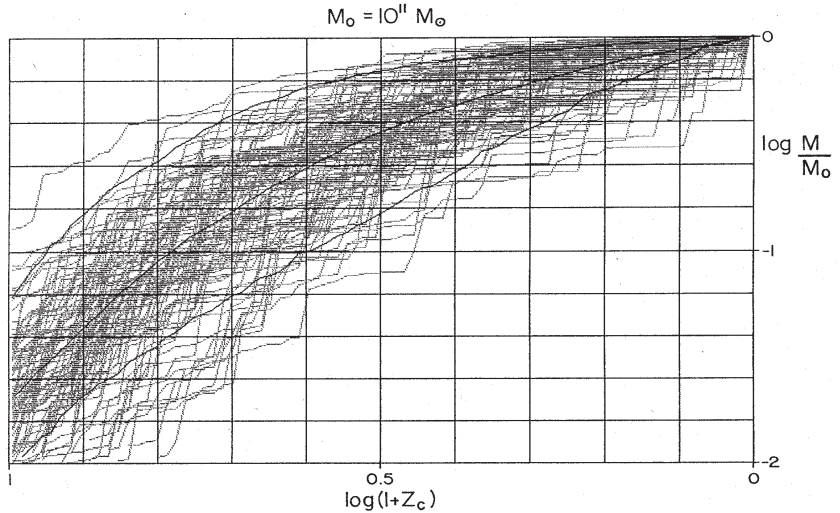}
\caption{Statistical realizations of possible mass growth histories 
for a 10$^{11}M_{\odot }$ dark halo. From top to bottom the low, average, and high accretion regimes are plotted (see text).}
\end{figure}

To study the virialization in more general cases (non
scale-invariant $MGHs$, nonradial orbits and $F(r)\neq const$) the
shell-crossing phases should be calculated. The crucial assumption in the
simplification of the problem is to consider the orbital period of the inner
shell much less than the time-scale of accretion of outer shells. In such
case the dynamics of the inner shell admits an adiabatic invariant, that for
radial orbits yields

\begin{equation}
M(<r_{ap})\propto r_{ap}^{-1}  \label{eq6}
\end{equation}
(Gunn 1977). $r_{ap}$ is the apapsis radius, where a particle spends most of
its orbital time. It is just the secular contraction of $r_{ap},$ by the
penetration of outer shells, that should be calculated . The procedure of Gunn 1977 and Filmore \& Goldreich 1984 consists of evaluating
the change $F^{-1}$of the enclosed mass within a given shell due to the
infall of outer shells, and then one uses relation (6). The mass
contained within a given apapsis $r_p$ consists of two contributions: the mass
of all particles with apapsis radius less than $r_{ap}$, and the mass of
outer infalling particles which spend some fraction of their orbital time
within $r_{ap}$. The evaluation of the second contribution is the crucial
step (Zaroubi \&\ Hoffman 1993). We use a statistical approach to the
problem, and generalize the cases of Zaroubi \&\ Hoffman for nonradial
orbits and arbitrary profiles ($MGHs$ in our case). An iterative procedure is
used to find $F^{-1}$. Typically the convergence is fast. Thus, given the $%
MGH,$ we calculate semi-analytically the structure of the collapsed
configuration. The degree of orbital nonradiality measured through $E\equiv 
\frac{\min or-axis}{mayor-axis},$ is a free parameter which affects the
central regions of the virialized halo. If $E<<1$ (radial case) the halo
tends to be coreless (see the rotation curves in Fig.5 )

\section{The mass growth histories}

 The statistical properties of the primordial density fluctuation field
depend on the physical mechanism in the very early universe that generated
the fluctuations. It is widely accepted (and as a {\em first}
approximation seems to work) possibility the inflation produced
a Gaussian random field with a near scale-invariant power spectrum. Since
the processed power spectrum after considering the damping and stagnation
processes is such that small scales have more density contrast, the
evolution of fluctuations will be hierarchical, from small to large
scales. Some statistics were worked out for this general case to
predict the development of the cosmic structure. An interesting step was the estimation of the conditional probability of finding a
collapsed object of mass $M_2$ at time $z_2$ provided it is embedded in a
larger collapsed object $M_1$ at an earlier time $z_1$ (Bower 1991, Bond $et$ $%
al.$ 1991, Lacey \&\ Cole 1993). This allows one to roughly follow the $MGHs$
back in time for a fixed mass at the present epoch. Note that because the
mass of the given object is forced to collapse today, we are already
selecting a special (isolated) region of the field. As was
pointed-out in the introduction, disk galaxies probably emerged from regions
of the field where the signal greatly dominates the noise, that is,
their $MGHs$ are nearer to a coherent accretion process, rather than a
chaotic and discontinuous aggregation of collapsed lumps.

As in Lacey \& Cole we generate Monte Carlo trajectories using the mentioned
probability distribution, and maintaining at every step the individuality of
the progenitor. The average of these special trajectories for a given
present-day mass is calculated and used as an initial condition of galaxy
formation. However, other
evolutionary trajectories are possible, covering a great variety of
disk galaxies with a given mass (the Hubble sequence). Thus, we
estimate the two, still significant, extrema by finding the deviations
from the most probable trajectory which constitutes roughly 10\% of all the
realizations and averaging them. Since the $MGHs$ are fixed to the present
mass, these extreme trajectories reflect the range of possible accretion
regimes of disk galaxies which for galactic masses oscillates between $%
\gamma \sim 0.2$ (low accretion regime) and $\gamma \sim 1$ (high accretion
regime). The average trajectory also slowly varies with mass ($\gamma $
rises with the mass). We used a standard cold dark matter model with power
spectrum normalized to $COBE$ (Stompor $et$ $al.$ 1995). Fig. 1 shows a
sample of 1000 halo histories (only 1 of every 10 trajectories are plotted)
for a dark halo of 10$^{11}M_{\odot }$ and the low, average, and high
accretion regime's cases.

\section{Results}

\begin{figure}
\vspace{8.0cm}
\includegraphics{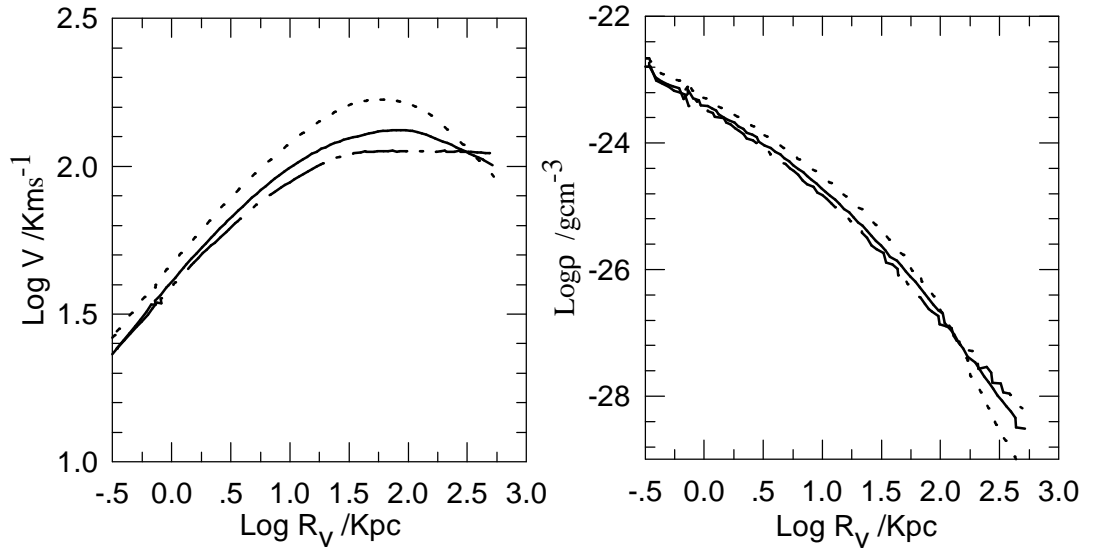} 
\caption{The circular velocities and density profiles of a 10$^{12}M_{\odot }$ dark halo. Dotted, solid, and dot.dashed lines correspond to the low, average, and high accretion regimes, respectively} 
\end{figure}

Figure 2 presents the rotation curves and density profiles of a 10$%
^{12}M_{\odot }$ halo corresponding to the low, average, and high accretion
rates. It is seen clearly that halos formed by fast collapse are more
concentrated than those formed by an extended collapse. In Figure 3 the
rotation curves for the average trajectory of 10$^{10}M_{\odot}$, 10$^{11}M_{\odot}, $ and 10$^{12}M_{\odot }$ halos are plotted. Low mass systems turn out to
be a little more concentrated than high mass ones (see also Navarro $et$ $al.$ 1996 who calculated the structure of dark halos using N-body simulations). The combination of these
results suggests that if the structural properties of dark halos are
important in establishing the Hubble sequence's properties of visible
galaxies, then this sequence is defined not only by one parameter, $i.e$ the mass, but at least by one more, the accretion regime. In $FAH$ we
show that in fact the accretion regime is also important in defining the star formation histories of galactic disks.

The relation between mass and maximum circular velocity (the
''cosmological'' Tully-Fisher relation $T-F$) we obtain from the average
trajectories is $M\propto V^{3.1}.$ Properly accounting for the
deviations around these trajectories the $r.m.s.$ scatter in the cosmological 
$T-F$ drops slowly for high mass halos, with a rough average of $\Delta
V/V\simeq 0.25$ corresponding to $\sigma \simeq 0.8mag.$ This is a high
value compared with the observational $T-F$ scatter. The evolutionary models
presented in $FAH$ give a scatter in B luminosity of $\sim 0.5mag.$
Some mechanism intrinsic to galactic evolution or related to the formation
of the disk could have reduced this scatter.

In order to obtain rotation curves resembling those of the observed
galaxies, it is necessary to consider the further dissipative collapse of
baryonic matter and its gravitational pull on the dark halo. This effect was
calculated employing a simple analytical model (Flores $et$ $al.$ 1993)
based on the adiabatic invariant formalism (see Firmani $et$ $al$. 1996).
The disk is built up as explained in $FAH.$ The effect is considerable in
the central halo regions. Figure 4 shows the rotation curve decomposition for
a 10$^{12}M_{\odot }$ galaxy. The halo rotation curves before and after the
disk formation are plotted. The  galactic rotation curves in all cases are nearly flat (see also figure 3 in $FAH$). However, going into more detail, for the range of masses analized here (10$^{10}M_{\odot}$-10$^{12}M_{\odot}$) the rotation curves within the optical radius show a very slow variation in shape, from gently rising in low mass galaxies to declining in more massive galaxies. Based on a large data set Persic \& Salucci 1995, and Persic $et$ $al$. 1996 have established  a similar trend, but more pronounced than the results of our models.

\begin{figure}[t]
\vspace{6.2cm}
\includegraphics{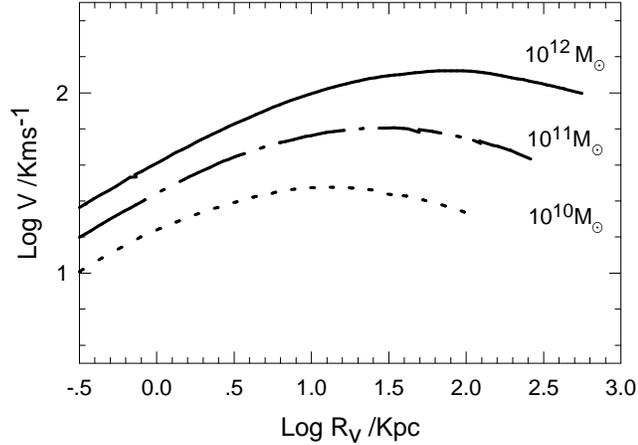}
\caption{The circular velocities of dark halos of 10$^{10}M_{\odot
}, $ 10$^{11}M_{\odot}, $ and 10$^{12}M_{\odot }$ in the case of average accretion regime. Low mass halos reach their maximum circular velocity at smaller radii w.r.t. the total virial radius than higher mass halos}
\end{figure}

\begin{figure}
\vspace{5.7cm}
\includegraphics{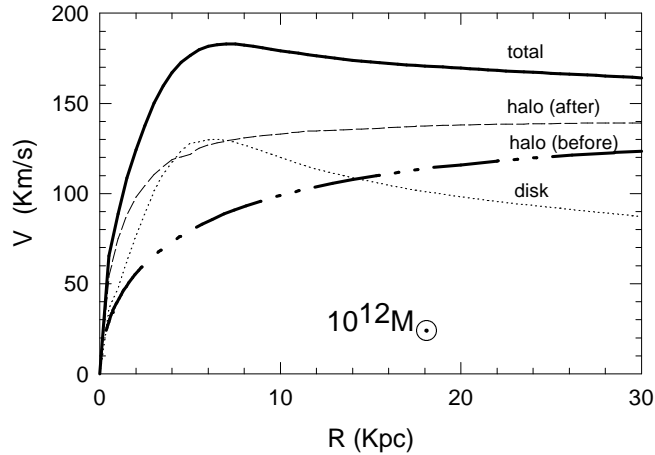}
\caption{Rotation curve decomposition for a 10$^{12}M_{\odot }$ galaxy and for the average accretion regime case}
\end{figure}

\section{Discusion and conclusions}

In section 2 it was mentioned that the degree of nonradiality $E$ is a free
parameter in our calculations. $E$ was fixed in such a way that the galactic
halos ($1$) have a core, and $(2)$ the disk rotation curve be of the order
of the halo rotation curve. It is well known that dwarf galaxies are
dominated by dark matter even in their central regions. So their rotation
curves are a good tracer of the dark halo. As Moore 1994 and Flores $et$ $%
al. $ 1995 pointed out, the observed rotation curves of dwarf galaxies show
the existence of a pronounced core. On the other hand, standard rotation
curve decompositions give a dominant contribution of the disk over that of
the dark halo in the center of high luminosity (normal) galaxies. For comparison we reconstructed the rotation curves obtained by N-body
simulations in Navarro $et$ $al.$ 1996 using their fitting formula (a $COBE$
normalization is used here). Figure 5 shows the comparison of these rotation
curves with the one obtained here. If we fix $E$ to lower values the
agreement is good, although the observational criteria ($1$) and $(2)$ will
not be satisfied. This discrepancy is atennuated for low-$\Omega$ CDM models (Navarro 1996). It could be also suggesting a more critical
analysis of the underpinning of the structure formation and evolution
theory, namely the random-phases hypothesis. Small scales, due to the
shape of the power spectrum, are more sensitive to this hypothesis.

 It is important to note that Persic  $et$ $al.$ 1996 have found that the degree of dominance of disk over halo in the central regions of galaxies is an increasing monotonic function of luminosity. This could be related to the result mentioned above that low mass dark halos are systematically more concentrated than their high mass counterparts. On the other hand, it was stated that for every mass, due to the different possible $MGHs$,  there is a sequence of dark structures ---a factor which could introduce a large dispersion in the relation between the galactic mass (but still not the disk luminosity) and  the disk$/$halo mass ratio. This question will be treated in detail in a future paper.

\begin{figure}
\vspace{6.5cm}
\includegraphics{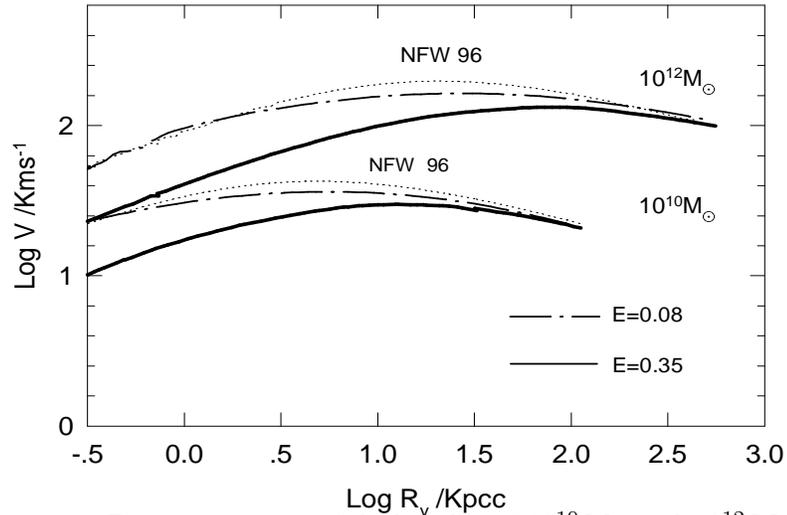}
\caption{Rotation curves of dark halos of 10$^{10}M_{\odot }$ and  10$^{12}M_{\odot }$ for the average accretion regime case and for two different degrees of orbital nonradiality (shown in the figure). The dotted lines correspond to the fitting formula of N-body simulations (Navarro $et$ $al.$ 1996)}
\end{figure}

 Throughout the present work it was demonstrated that the amplitude and shape of
galactic rotation curves for a given mass and cosmological model depends on
the accretion regime, the level of substructure around the peaks, and the
further collapse of baryonic matter. The different accretion regimes
generated from initial cosmological conditions establish a range of
structural halo properties that could be important in understanding the
origin of the Hubble sequence.

\acknowledgments
We thank Alberto Garc\'{\i}a  for thechnical help in preparing some of the figures, and Stan Kurtz, and Pedro Col\'{\i}n for their critical reading of the manuscript. V.A-R. received partial financial support from the PADEP-UNAM program. He akcnowledges a fellowship from CONACyT under the iberoamerican program MUTIS.

\end{document}